\documentclass[ALICE,manyauthors,onecolumn]{cernphprep}
\usepackage[comma,square,numbers,sort&compress]{natbib}
\usepackage{hyperref}
\usepackage{lineno}
\usepackage{multirow}
\usepackage{amsmath}
\usepackage{siunitx}
\usepackage{xcolor}
\usepackage{wasysym}
\usepackage{placeins}
\usepackage{bm}
\usepackage[T1]{fontenc}
\usepackage{enumitem}
\usepackage{breakurl}
\usepackage[Symbol]{upgreek}

\DeclareSIUnit\fm{\femto\meter}
\DeclareSIUnit\eVc{\eV\per\clight}
\DeclareSIUnit\keVc{\keV\per\clight}
\DeclareSIUnit\MeVc{\MeV\per\clight}
\DeclareSIUnit\MeVcc{\MeV\per\clight\squared}
\DeclareSIUnit\GeVc{\GeV\per\clight}
\DeclareSIUnit\GeVcc{\GeV\per\clight\squared}
\DeclareSIUnit\clight{\text{\ensuremath{c}}}
\DeclareSIUnit[number-unit-product = ]\percent{\char`\%}

\newcommand{\onethree}     {$\sqrt{s}~=~13$~Te\kern-.1emV\xspace}

\newcommand{\pt}{\ensuremath{p_{\rm T}}\,} 
\newcommand{\mt}{\ensuremath{m_{\rm T}}\,} 
\newcommand{\kt}{\ensuremath{k_{\rm T}}\,}

\newcommand{\La}{\ensuremath{\Lambda}\,}
\newcommand{\aLa}{\ensuremath{\overline{\Lambda}}\,}
\newcommand{\sig}{\ensuremath{\Sigma}\,}
\newcommand{\sz}{\ensuremath{\Sigma^{0}}\,}
\newcommand{\xim}{\ensuremath{\Xi^{-}}\xspace}
\newcommand{\xip}{\ensuremath{\overline{\Xi}^{+}}\xspace}

\newcommand{\omm}{\ensuremath{\Omega^{-}}\xspace}
\newcommand{\omp}{\ensuremath{\overline{\Omega}^{+}}\xspace}

\newcommand{\pp}{pp\xspace}
\newcommand{\ppNoSpace}{\ensuremath{\mathrm {p\kern-0.05em p}}}
\newcommand{\pPb}{\ensuremath{\mbox{p--Pb}}\,}
\newcommand{\AuAu}{\mbox{Au--Au}\xspace}

\newcommand{\NN}{\ensuremath{\mbox{N--N}}\,}
\newcommand{\YN}{\ensuremath{\mbox{Y--N}}\,}

\newcommand{\YY}{\ensuremath{\mbox{Y--Y}}\,}
\newcommand{\YYbound}{\ensuremath{\mbox{Y--Y}}\,}
\newcommand{\YNbound}{\ensuremath{\mbox{Y--N}}\,}
\newcommand{\LN}{\ensuremath{\mbox{$\Lambda$--N}}~}

\newcommand{\pP}{\ensuremath{\mbox{p--p}}~}

\newcommand{\pXim}{\ensuremath{\mbox{p--$\Xi^{-}$}}\xspace}
\newcommand{\pXip}{\ensuremath{\overline{\mathrm{p}}$\mbox{--}$\xip}\xspace}

\newcommand{\pOmbound}{\ensuremath{\mbox{p--$\Omega^{-}$}}\xspace}
\newcommand{\pOm}{\ensuremath{\mbox{p--$\Omega^{-}$}}\xspace}
\newcommand{\pOp}{\ensuremath{\overline{\mathrm{p}}$\mbox{--}$\omp}\xspace}

\newcommand{\LL}{\ensuremath{\mbox{$\Lambda$--$\Lambda$}}~}

\newcommand{\ks}           {\ensuremath{k^*}\xspace}
\newcommand{\rs}           {\ensuremath{r^*}\xspace}

\newcommand{\MeVc}{\ensuremath{\mathrm{MeV}\kern-0.05em/\kern-0.02em \textit{c}}~}
\newcommand{\GeVc}{\ensuremath{\mathrm{GeV}\kern-0.05em/\kern-0.02em \textit{c}}~}
\newcommand{\GeVcSq}{\ensuremath{\mathrm{GeV}\kern-0.05em/\kern-0.02em \textit{c}^2}~}
\newcommand{\MeVcSq}{\ensuremath{\mathrm{MeV}\kern-0.05em/\kern-0.02em \textit{c}^2}~}

\newcommand{\HALQCD}{HAL QCD\xspace}

\newcommand{\radiuspXi}{\ensuremath{1.02\pm0.05}\,}\newcommand{\radiuspOmega}{\ensuremath{0.95\pm0.06}\,}
\newcommand{\fakeXi}{\ensuremath{8}\,}
\newcommand{\fakeOmega}{\ensuremath{5}\,}
\newcommand{\events}{\ensuremath{1\times10^9}\,}

\newcolumntype{b}{X}
\newcolumntype{s}{>{\hsize=.5\hsize}X}

\begin{document}

\begin{titlepage}
\PHyear{2020}
\PHnumber{091}      \PHdate{20 May}  \title{Unveiling the strong interaction among hadrons at the LHC}
\ShortTitle{Unveiling the strong interaction among hadrons at the LHC}

\Collaboration{ALICE Collaboration\thanks{See Appendix~\ref{app:collab} for the list of Collaboration members}}
\ShortAuthor{ALICE Collaboration} 

\begin{abstract}
One of the key challenges for nuclear physics today is to understand from first principles the effective interaction between hadrons with different quark content.
First successes have been achieved using techniques that solve the dynamics of quarks and gluons on discrete space-time lattices~\cite{Beane:2013br,Epelbaum:2009pd}.
Experimentally, the dynamics of the strong interaction have been studied by scattering hadrons off each other.
Such scattering experiments are difficult or impossible for unstable hadrons~\cite{Eisele:1971mk,Alexander:1969cx,SechiZorn:1969hk,Muller:1972cq} and so high-quality measurements exist only for hadrons containing up and down quarks ~\cite{PhysRevC.47.761}.
Here we demonstrate that measuring correlations in the momentum space between hadron pairs~\cite{FemtoRun1,Acharya:2019bsa,FemtoLambdaLambda,FemtopXi,Acharya:2019kqn} produced in ultrarelativistic proton--proton collisions at the CERN Large Hadron Collider (LHC) provides a precise method with which to obtain the missing information on the interaction dynamics between any pair of unstable hadrons. 
Specifically, we discuss the case of the interaction of baryons containing strange quarks (hyperons).
We demonstrate how, using precision measurements of \pOm baryon correlations, the effect of the strong interaction for this hadron--hadron pair can be studied with precision similar to, and compared with, predictions from lattice calculations~\cite{Iritani:2018sra, Sasaki:2019qnh}.
The large number of hyperons identified in proton--proton collisions at the LHC, together with an accurate modelling~\cite{Acharya:2020dfb} of the small (approximately one femtometre) inter-particle distance and exact predictions for the correlation functions, enables a detailed determination of the short-range part of the nucleon-hyperon interaction.

\end{abstract}
\end{titlepage}
\setcounter{page}{2}

\section*{Introduction}
Baryons are composite objects formed by three valence quarks bound together by means of the strong interaction mediated through the emission and absorption of gluons.
Between baryons, the strong interaction leads to a residual force and the most common example is the effective strong force among nucleons (N)--baryons composed of up (\textit{u}) and down (\textit{d}) quarks: proton (p) = \textit{uud} and neutron (n) = \textit{ddu}.
This force is responsible for the existence of a neutron--proton bound state, the deuteron, and manifests itself in scattering experiments~\cite{PhysRevC.47.761} and through the
existence of atomic nuclei.
So far, our understanding of the nucleon--nucleon strong interaction relies heavily on effective theories~\cite{RevModPhys.81.1773}, where the degrees of freedom are nucleons.
These effective theories are constrained by scattering measurements and are successfully used in the description of nuclear properties~\cite{Hebeler:2015hla,PhysRevLett.118.152503}.
 
 The fundamental theory of the strong interaction is quantum chromodynamics (QCD), in which quarks and gluons are the degrees of freedom. One of the current challenges in nuclear physics is to calculate the strong interaction among hadrons starting from first principles. Perturbative techniques are used to calculate strong-interaction phenomena in high-energy collisions with a level of precision of a few per cent~\cite{Klijnsma:2017eqp}.
For baryon--baryon interactions at low energy such techniques cannot be employed; however, numerical solutions on a finite space-time lattice have been used to calculate scattering parameters among nucleons and the properties of light nuclei \cite{Beane:2013br,Epelbaum:2009pd}.
 Such approaches are still limited: they do not yet reproduce the properties of the deuteron~\cite{Orginos:2015aya} and do not predict physical values for the masses of light hadrons \cite{Wagman:2017tmp}.

Baryons containing strange (\textit{s}) quarks, exclusively or combined with \textit{u} and \textit{d} quarks, are called hyperons (Y) and are denoted by uppercase Greek letters: \La = \textit{uds}, \sz = \textit{uds}, \xim = \textit{dss}, \omm = \textit{sss}. 
Experimentally, little is known about \YN and \YY interactions, but recently, major steps forward in their understanding have been made using lattice QCD approaches~\cite{Iritani:2018sra, Sasaki:2019qnh,Hatsuda:2018nes}.
The predictions available for hyperons are characterized by smaller uncertainties because the lattice calculation becomes more stable for quarks with larger mass, such as the \textit{s} quark.
In particular, robust results are obtained for interactions involving the heaviest hyperons, such as $\Xi$ and $\Omega$, and precise measurements of the \pXim and \pOm  interactions are instrumental in validating these calculations. 
From an experimental point of view, the existence of nuclei in which a nucleon is replaced by a hyperon (hypernuclei) demonstrates the presence of an attractive strong \LN interaction~\cite{Hashimoto:2006aw} and indicates the possibility of binding a \xim to a nucleus~\cite{Nakazawa:2015joa,Nagae:2017slp}.
A direct and more precise measurement of the \YN interaction requires scattering experiments,
which are particularly challenging to perform because hyperons are short-lived and travel only a few centimetres before decaying.
Previous experiments with $\Lambda$ and $\Sigma$ hyperons on proton targets~\cite{Eisele:1971mk,Alexander:1969cx,SechiZorn:1969hk} delivered results that were two orders of magnitude less precise than those for nucleons, and such experiments with $\Xi$ (ref. \cite{Muller:1972cq}) and $\Omega$ beams are even more challenging. 
The measurement of the \YN and \YY interactions has further important implications for  the possible formation of a \YNbound or \YYbound bound state.  
Although numerous theoretical predictions exist~\cite{Francis:2018qch,Jaffe:1976yi,PhysRevD.15.2547,PhysRevD.20.1633,Gongyo:2017fjb,Iritani:2018sra}, so far no clear evidence for any such bound states has been found,
despite many experimental searches~\cite{Adam:2015nca,Kim:2013vym,Chrien:1998yt,Yoon:2007aq,Ahn:1998fj}. 

Additionally, a precise knowledge of the \YN and \YY interactions has important consequences for the physics of neutron stars. Indeed, the structure of the innermost core of neutron stars is still completely unknown and hyperons could appear in such environments depending on the \YN and \YY interactions~\cite{Tolos:2020aln}. Real progress in this area calls for new experimental methods.

Studies of the \YN interaction via correlations have been pioneered by the HADES collaboration~\cite{Adamczewski-Musch:2016jlh}. Recently, the ALICE collaboration has demonstrated that \pp and \pPb collisions at the LHC are best suited to study the \NN and several \YN, \YY interactions precisely \cite{FemtoRun1,Acharya:2019bsa,FemtoLambdaLambda,FemtopXi,Acharya:2019kqn}.
Indeed, the collision energy and rate available at the LHC opens the phase space for an abundant production of any strange hadron~\cite{ALICE:2017jyt}, and the capabilities of the ALICE detector for particle identification and the momentum resolution--with values below \SI{1}{\percent} for transverse momentum $\pt<\SI{1}{\GeVc}$--facilitate the investigation of correlations in momentum space.
These correlations reflect the properties of the interaction and hence can be used to test theoretical predictions by  solving the Schr\"odinger equation for proton--hyperon collisions~\cite{Mihaylov:2018rva}.
A fundamental advantage of \pp and \pPb collisions at LHC energies is the fact that all hadrons originate from very small space-time volumes, with typical inter-hadron distances of about \SI{1}{\fm}.
These small distances are linked through the uncertainty principle to a large relative momentum range (up to \SI{200}{\MeVc}) for the baryon pair and enable us to test short-range interactions.
Additionally, detailed modelling of a common source for all produced baryons \cite{Acharya:2020dfb} allow us to determine accurately the source parameters.

Similar studies were carried out in ultrarelativistic \AuAu collisions at a centre-of-mass energy of \SI{200}{\GeV} per nucleon pair by the STAR collaboration for \LL~\cite{Adamczyk:2014vca,Morita:2014kza} and \pOm~\cite{STAR:2018uho} interactions.
This collision system leads to comparatively large particle emitting sources of 
\num{3}--\SI{5}{\fm}. The resulting relative momentum range
is below \SI{40}{\MeVc}, implying reduced sensitivity to interactions at distances shorter than \SI{1}{\fm}.  

In this work, we present a precision study of the most exotic among the proton--hyperon interactions, obtained via the \pOm correlation function in pp collisions at a centre-of-mass energy $\sqrt{s}=13$~Te\kern-.1emV\xspace at the LHC. 
The comparison of the measured correlation function with first-principle calculations ~\cite{Iritani:2018sra} and with a new precision measurement of the \pXim correlation in the same collision system
provides the first observation of the effect of the strong interaction for the \pOm pair.
The implications of the measured correlations for a possible \pOmbound bound state are also discussed.
These experimental results challenge the interpretation of the data in terms of lattice QCD as the precision of the data improves.

Our measurement opens a new chapter for experimental methods in hadron physics with the potential to pin down the strong interaction for all known proton--hyperon pairs.

 \section*{Analysis of the Correlation Function }
\begin{figure}[h]
  \centering
  \includegraphics[width=0.85\linewidth]{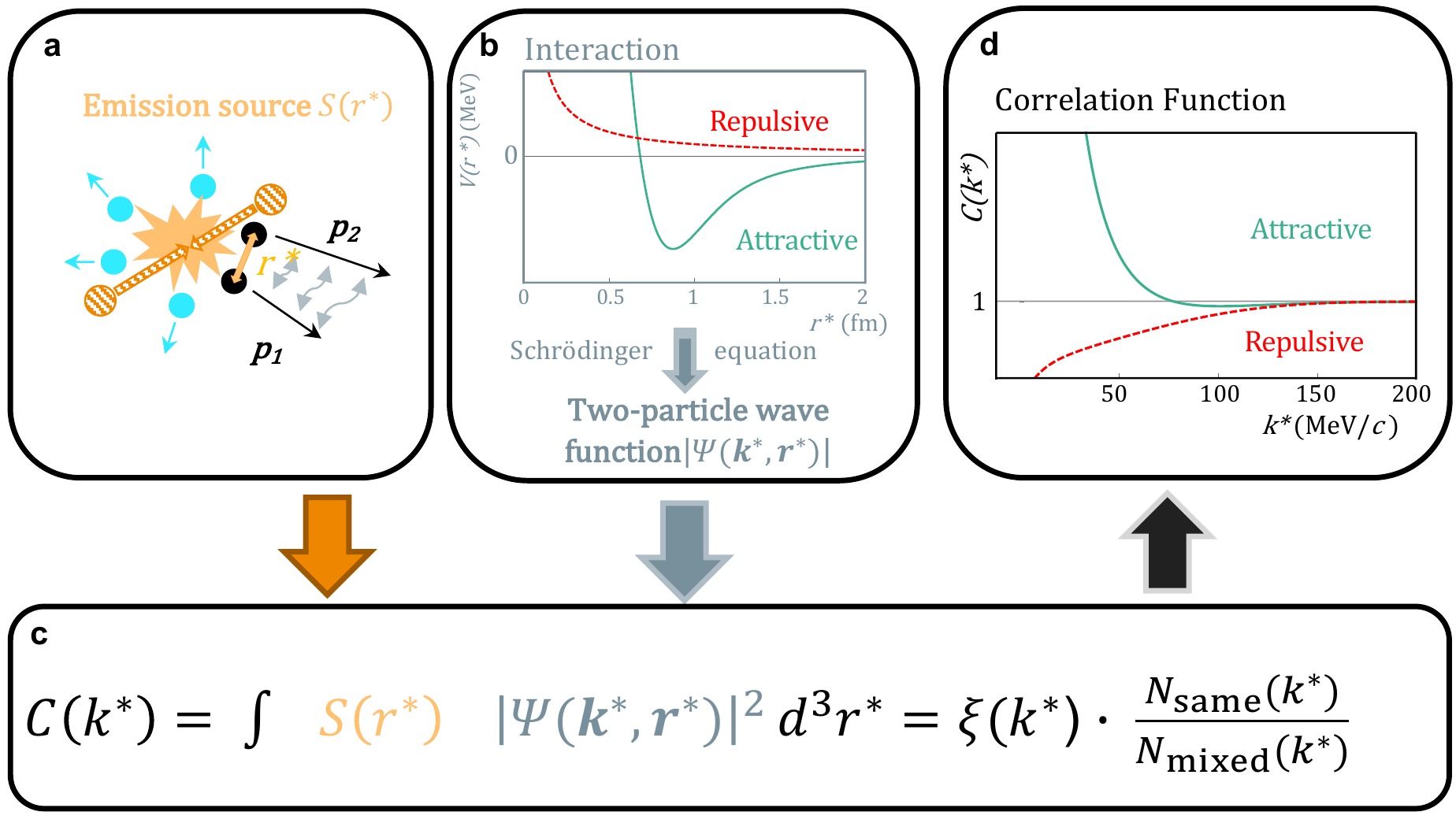}
  \caption{\textbf{Schematic representation of the correlation method.}
  \textbf{a}, A collision of two protons generates a particle source $S(\rs)$ from which a hadron--hadron pair with momenta $\bm{p}\bf{_1}$ and $\bm{p}\bf{_2}$ emerges at a relative distance $r^*$ and can undergo final-state interaction before being detected.
Consequently, the relative momentum \ks is either reduced or increased via an attractive or a repulsive interaction, respectively.
  \textbf{b}, Example of attractive (green) and repulsive (dotted red) interaction potentials, V(\rs), between two hadrons, as a function of their relative distance.
Given a certain potential, a non-relativistic Schr\"odinger equation is used to obtain the corresponding two-particle wave function, $\psi(\bm{k}^*,\bm{r}^*)$.
  \textbf{c}, The equation of the calculated (second term) and measured (third term) correlation function \textit{C}(\ks), where $N_{\mathrm{same}}(\ks)$ and $N_{\mathrm{mixed}}(\ks)$ represent the \ks distributions of hadron--hadron pairs produced in the same and in different collisions, respectively, and $\xi(\ks)$ denotes the corrections for experimental effects.
  \textbf{d}, Sketch of the resulting shape of \textit{C}(\ks).
The value of the correlation function is proportional to the interaction strength.
It is above unity for an attractive (green) potential, and between zero and unity for a repulsive (dotted red) potential.
  }
  \label{fig:Femtoscopy}
\end{figure}

Figure~\ref{fig:Femtoscopy} shows a schematic representation of the correlation method used in this analysis. 
The correlation function can be expressed theoretically~\cite{Pratt:1986cc,Lisa:2005dd} as 
$C(\ks)\,=\,\int d^3\rs S(\rs) \times|{\psi(\bm{k}^*,\bm{r}^*)}|^2$, 
where $\bm{k}^*$ and $\bm{r}^*$ are the relative momentum and the relative distance of the pair of interest.
$S(\rs)$ is the distribution of the distance $r^*=|\bm{r}^*|$ at which particles are emitted (defining the source size),
$\psi(\bm{k}^*,\bm{r}^*)$ represents the wave function of the relative motion for the pair of interest and $k^*=|\bm{k}^*|$ is the reduced relative momentum of the pair ($\ks\,=\,|\bm{p}\bf{^*_2}-\bm{p}\bf{^*_1}|$/2).
Given an interaction potential between two hadrons as a function of their relative distance,
a non-relativistic Schr\"odinger equation can be used~\cite{Mihaylov:2018rva} to obtain the corresponding wave function and hence also predict the expected correlation function.
The choice of a non-relativistic Schr\"odinger equation is motivated by the fact that the typical relative momenta relevant for the strong final-state interaction have a maximal value of \SI{200}{\MeVc}. 
Experimentally, this correlation function is computed as $C(\ks)\,=\,\xi(\ks)\frac{N_{\mathrm{same}}(\ks)}{N_{\mathrm{mixed}}(\ks)}$,
where $\xi(\ks)$ denotes the corrections for experimental effects,
$N_{\mathrm{same}}(\ks)$ is the number of pairs with a given \ks obtained by combining particles produced in the same collision (event), which constitute a sample of correlated pairs, and $N_{\mathrm{mixed}}(\ks)$ is the number of uncorrelated pairs with the same $\ks$, obtained by combining particles produced in different collisions (the so-called mixed-event technique).
Fig.~\ref{fig:Femtoscopy}, panel $\textbf{d}$, shows how an attractive or repulsive interaction is mapped into the correlation function. For an attractive interaction the magnitude of the correlation function will be above unity for small values of \ks, whereas for a repulsive interaction it will be between zero and unity.
In the former case, the presence of a bound state would create a depletion of the correlation function with a depth increasing with increasing binding energy.

Correlations can occur in nature from quantum mechanical interference, resonances, conservation laws or final-state interactions. Here, it is the final-state interactions that contribute predominantly at low relative momentum; in this work we focus on the strong and Coulomb interactions in pairs composed of a proton and either a \xim or a \omm hyperon. 

\begin{figure*}[hbt]
  \centering
  \includegraphics[width=0.55\linewidth]{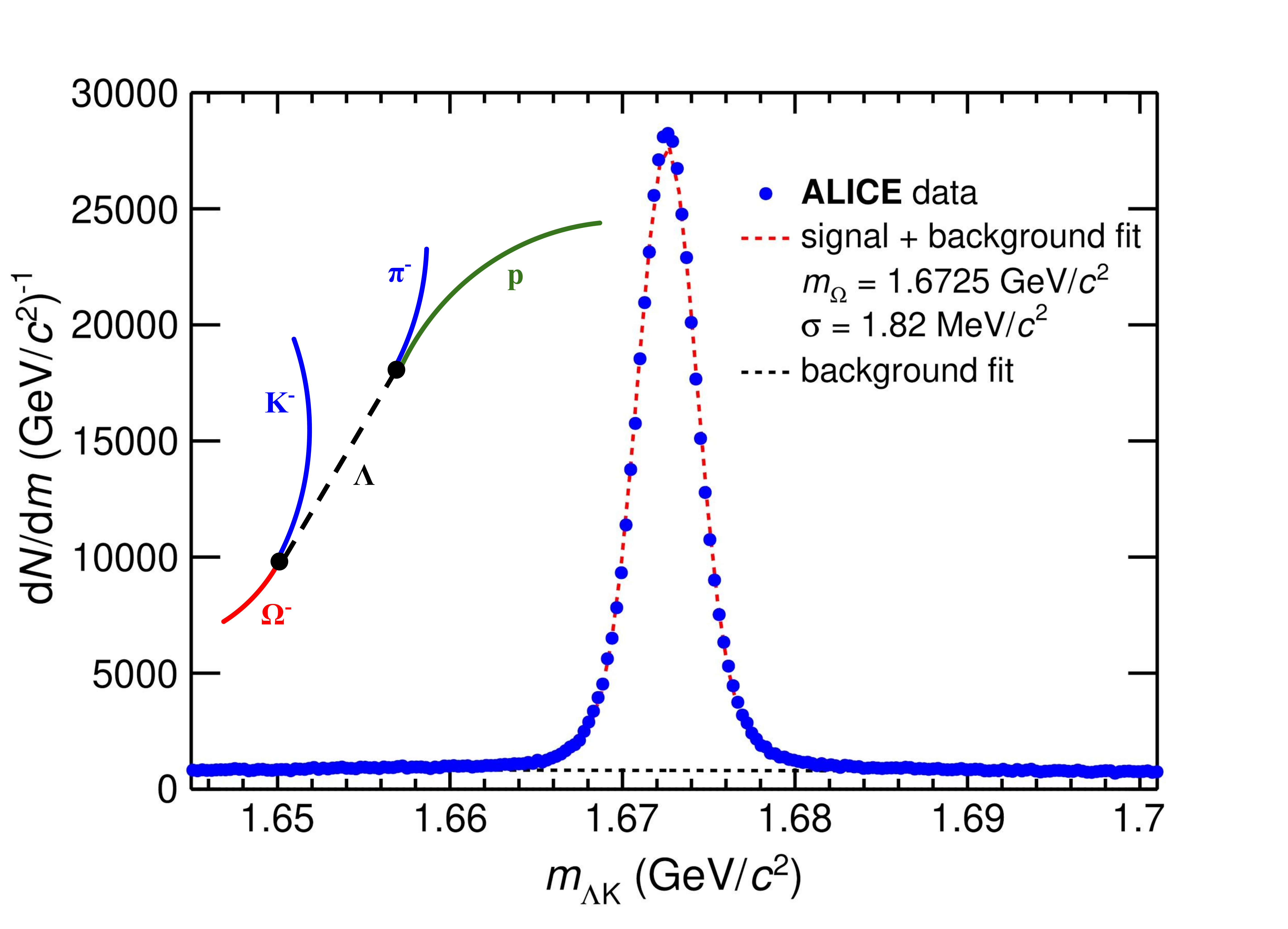}
  \hfill
\caption{\textbf{Reconstruction of the \omm and \omp signals.} Sketch of the weak decay of \omm into a $\Lambda$ and a K$^{-}$,
and measured invariant mass distribution (blue points) of $\Lambda$K$^{-}$ and $\overline{\Lambda}$K$^{+}$ combinations.
The dotted red line represents the fit to the data including signal and background,
and the black dotted line the background alone.
The contamination from misidentification is $\leq$ \fakeOmega\si{\percent}. }
 \label{fig:Minv}
\end{figure*}

Protons do not decay and can hence be directly identified within the ALICE detector, but \xim and \omm baryons are detected through their weak decays, 
$\xim\rightarrow \La + \uppi^{-}$ and $\omm\rightarrow \La + $K$^{-}$.
The identification and momentum measurement of protons, \xim, \omm and their respective antiparticles are described in \nameref{sec:Methods}.
Figure~\ref{fig:Minv} shows a sketch of the \omm decay and the invariant mass distribution of the $\Lambda$K$^{-}$ and $\overline{\Lambda}$K$^{+}$ pairs.
The clear peak corresponding to the rare \omm and \omp baryons demonstrates the excellent identification capability, which is the key ingredient for this measurement.
The contamination from misidentification is $\leq$ \fakeOmega\si{\percent}. For the \xim (\xip) baryon the misidentification amounts to \fakeXi\si{\percent}~\cite{FemtopXi}.

 Once the p, $\omm$ and $\xim$ candidates and charge conjugates are selected and their 3-momenta measured, the correlation functions can be built. 
Since we assume that the same interaction governs baryon--baryon and antibaryon--antibaryon pairs~\cite{FemtoRun1},
we consider in the following the direct sum ($\oplus$) of particles and antiparticles (\pXim$ \oplus$ \pXip $\equiv$ \pXim and \pOm $\oplus$ \pOp $\equiv$ \pOm).
The determination of the correction $\xi(\ks)$ 
and the evaluation of the systematic uncertainties are described in \nameref{sec:Methods}.

\section*{Comparison of the \pXim and \pOm interactions}
\begin{figure}[ht]
\centering
  \includegraphics[width = 0.5\textwidth]{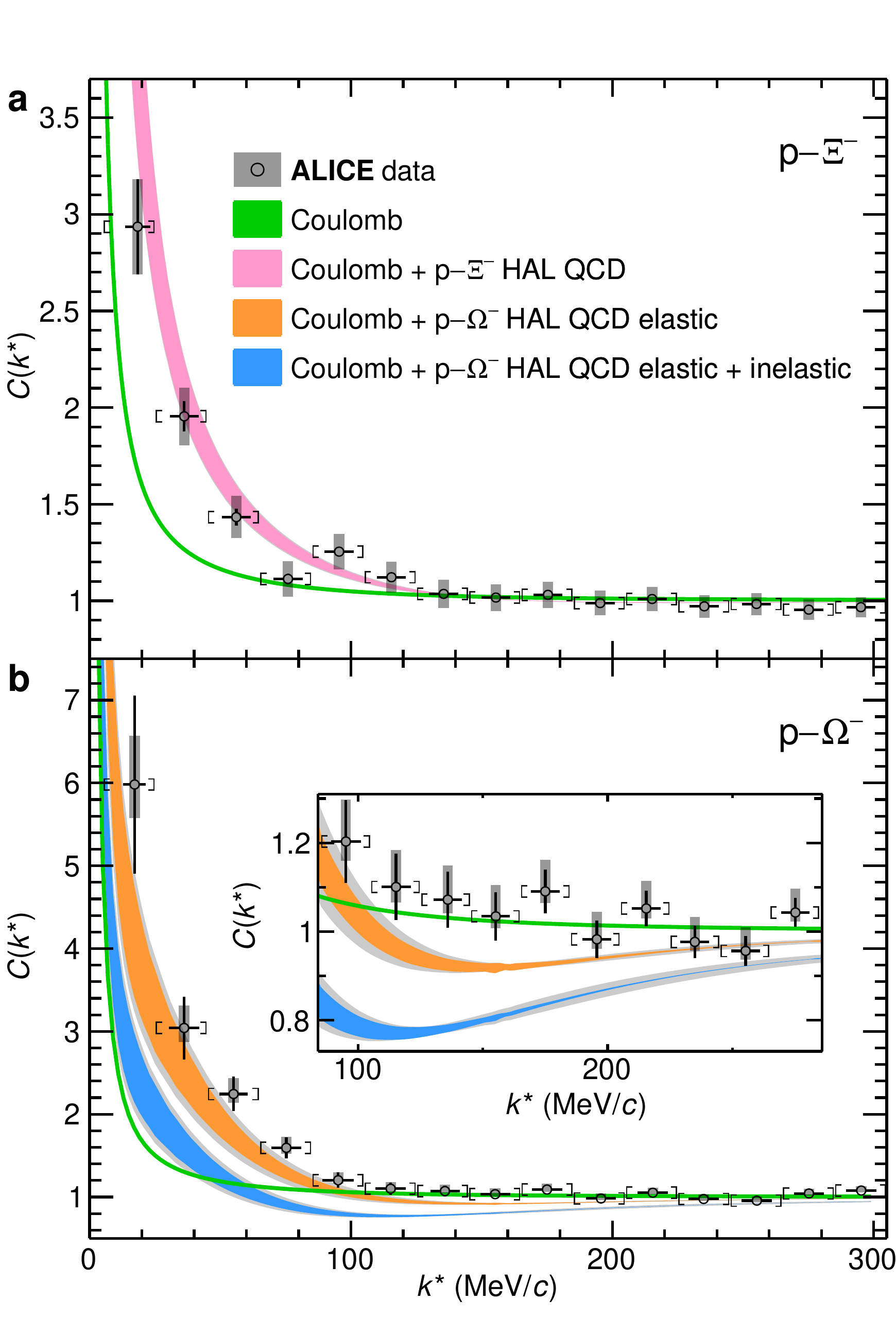}
\caption{\textbf{Experimental \pXim and \pOm correlation functions. }
\textbf{a}, \textbf{b}, Measured \pXim (\textbf{a}) and \pOm (\textbf{b}) correlation functions in high multiplicity \pp collisions at $\sqrt{s}=13$~TeV\xspace.
The experimental data are shown as black symbols.
The black vertical bars and the grey boxes represent the statistical and systematic uncertainties.
The square brackets show the bin width and the horizontal black lines represent the statistical uncertainty in the determination of the mean \ks for each bin.
The measurements are compared with theoretical predictions, shown as coloured bands, that assume either Coulomb or Coulomb + strong \HALQCD interactions.
For the \pOm system the orange band represents the prediction considering only the elastic contributions and the blue band represents the prediction considering both elastic and inelastic contributions.
The width of the curves including \HALQCD predictions represents the uncertainty associated with the calculation (see \nameref{sec:Methods} section 'Corrections of the correlation function' for details) and the grey shaded band represents, in addition, the uncertainties associated with the determination of the source radius.
The width of the Coulomb curves represents only the uncertainty associated with the source radius.
The considered radius values are \radiuspXi \si{\fm} for \pXim and \radiuspOmega \si{\fm} for \pOm pairs, respectively. 
The inset in \textbf{b} shows an expanded view of the \pOm correlation
function for \textit{C}(\ks) close to unity.
For more details see text.
} 
 \label{fig:CF}
\end{figure}

The obtained correlation functions are shown Fig.~\ref{fig:CF}, panels $\textbf{a}$ and $\textbf{b}$, for the \pXim and \pOm pairs, respectively, along with the statistical and systematic uncertainties. 
The fact that both correlations are well above unity implies the presence of an attractive interaction for both systems.
For opposite-charge pairs, as considered here, the Coulomb interaction is attractive and its effect on the correlation function is illustrated by the green curves in both panels of Fig.~\ref{fig:CF}.
These curves have been obtained by solving the Schr\"odinger equation for \pXim and \pOm pairs using the Correlation Analysis Tool using the Schr\"odinger equation (CATS) equation solver~\cite{Mihaylov:2018rva}, considering only the Coulomb interaction and assuming that the shape of the source follows a Gaussian distribution with a width equal to \radiuspXi \si{\fm} for the \pXim system and to \radiuspOmega \si{\fm} for the \pOm system, respectively.
The source-size values have been determined via an independent analysis of \pP correlations~\cite{Acharya:2020dfb}, where modifications of the source distribution due to strong decays of short-lived resonances are taken into account, and the source size is determined as a function of the transverse mass \mt of the pair, as described in \nameref{sec:Methods}. The average \mt of the \pXim and \pOm pairs are \SI{1.9}{\GeVc} and \SI{2.2}{\GeVc}, respectively.
The difference in size between the source of the \pXim and \pOm pairs might reflect the contribution of collective effects such as (an)isotropic flow.
The width of the green curves in Fig.~\ref{fig:CF} reflects the quoted uncertainty of the measured source radius. 
The correlations obtained, accounting only for the Coulomb interaction, considerably underestimate the strength of both measured correlations. This implies, in both cases, that an attractive interaction exists and exceeds the strength of the Coulomb interaction. 

\begin{figure*}[ht]
  \centering
   \includegraphics[width=0.60\linewidth]{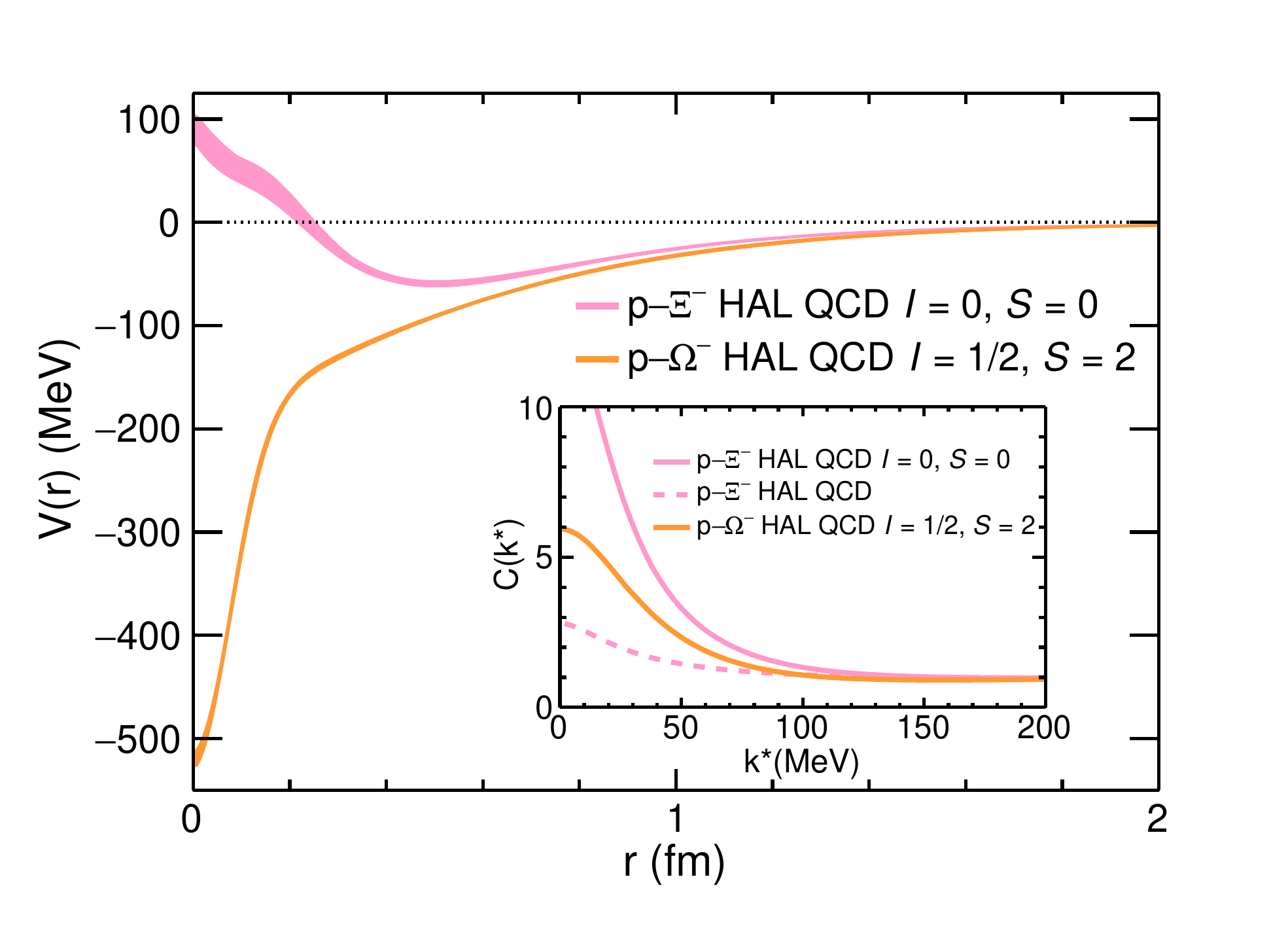}
  \hfill
\caption{\textbf{ Potentials for the \pXim and \pOm interactions.} \pXim (pink) and \pOm (orange) interaction potentials as a function of the pair distance predicted by the \HALQCD collaboration~\cite{Iritani:2018sra,Sasaki:2019qnh}. 
Only the most attractive component, isospin $I$~=~0 and spin $S$ = 0, is shown for \pXim. For the \pOm interaction the $I$~=~1/2 and spin $S$~=~2 component is shown.
The widths of the curves correspond to the uncertainties (see \nameref{sec:Methods} section 'Corrections of the correlation function' for details) associated with the calculations. 
The inset shows the correlation functions obtained 
using the HAL QCD strong interaction potentials
for: (i) the channel \pXim with isospin $I$~=~0 and spin $S$ = 0 , (ii) the channel \pXim including all allowed spin and isospin combinations (dashed pink), and (iii) the channel \pOm with isospin $I$~=~1/2 and spin $S$ = 2. For details see text.
}
 \label{fig:pot}
\end{figure*}

To discuss the comparison of the experimental data with the predictions from lattice QCD, it is useful to first focus on the distinct characteristics of the \pXim and \pOm interactions.
Figure~\ref{fig:pot} shows the radial shapes obtained for the strong-interaction potentials calculated from first principles by the \HALQCD (Hadrons to Atomic nuclei from Lattice QCD) collaboration for the \pXim (ref.~\cite{Sasaki:2019qnh}) and the \pOm systems~\cite{Iritani:2018sra}, see \nameref{sec:Methods} for details.
Only the most attractive (isospin $I$ = 0 and spin $S$ = 0) of the four components~\cite{Sasaki:2019qnh} of the \pXim interaction and the isospin $I$ = 1/2 and spin $S$ = 2 component of the \pOm interaction are shown.
Aside from an attractive component, we see that the interaction contains also a repulsive core starting at very small distances, below \SI{0.2}{\fm}.
For the \pOm system no repulsive core is visible and the interaction is purely attractive.
This very attractive interaction can accommodate a \pOmbound bound state, with a binding energy of about \SI{2.5}{\MeV}, considering the Coulomb and strong forces~\cite{Iritani:2018sra}.
The \pXim and \pOm interaction potentials look very similar to each other above a distance of \SI{1}{\fm}.
This behaviour is not observed in phenomenological models that engage the exchange of heavy mesons and predict a quicker fall off of the potentials \cite{Haidenbauer:2019utu}.

The inset of Fig. \ref{fig:pot} shows the correlation functions obtained 
using the HAL QCD strong interaction potentials
for: (i) the channel \pXim with isospin $I$~=~0 and spin $S$ = 0, (ii) the channel \pXim including all allowed spin and isospin combinations, and (iii) the channel \pOm  with isospin $I$~=~1/2 and spin $S$ = 2.
The correlation functions are computed using the experimental values for the \pXim and \pOm source-size. 
 Despite the fact that the strong \pOm potential is more attractive than the \pXim $I$~=~0 and $S$ = 0 potential, the resulting correlation function is lower. This is due to the presence of the bound state in the \pOm case \cite{Morita:2016auo}.
If we consider all four isospin and spin components of the \pXim interaction \cite{FemtopXi} 
the prediction for the global \pXim correlation function is lower than that for \pOm.
Experimentally, as shown in Fig.~\ref{fig:CF}, the less attractive strong \pXim interaction translates into a correlation function that reaches values of \num{3} in comparison with the much higher values of up to \num{6} that are visible for the \pOm correlation.
The theoretical predictions shown in Fig.~\ref{fig:CF} also include the effect of the Coulomb interaction.

Regarding the \pXim interaction, it should be considered that strangeness-rearrangement processes can occur, such as $\rm{p}\xim\rightarrow \La\La,\,\sig\sig,\,\La\sig$.
This means that the inverse processes (for example, $\La\La \rightarrow \rm{p}\xim$) can also occur and modify the \pXim correlation function.
These contributions are accounted for within lattice calculations by exploiting the well known quark symmetries~\cite{Sasaki:2019qnh} and are found to be very small.
Moreover, the ALICE collaboration measured the \LL correlation in \pp and \pPb collisions~\cite{FemtoLambdaLambda} and good agreement with the shallow interaction predicted by the \HALQCD collaboration was found.

The resulting prediction for the correlation function, obtained by solving the Schr\"odinger equation for the single \pXim channel including the \HALQCD strong and Coulomb interactions,
is shown in Fig.~\ref{fig:CF}, panel \textbf{a}.
The first measurement of the \pXim interaction using \pPb collisions~\cite{FemtopXi} showed a qualitative agreement to lattice QCD predictions.
The improved precision of the data 
in the current analysis of pp collisions is also in agreement with calculations that include both the \HALQCD and Coulomb interactions.

\section*{Detailed Study of the \pOm Correlation}
Concerning the \pOm interaction, strangeness-rearrangement processes can also occur~\cite{Morita:2019rph}, such as $ \rm{p}\omm \rightarrow \Xi \La,\, \Xi\sig$.
Such processes might affect the \pOm interaction in a different way depending on the
relative orientation of the 
total spin and angular momentum of the pair.
Since the proton has $J_p = 1/2$ and the $\Omega$ has $J_{\Omega} = 3/2$ and the orbital angular momentum $L$ can be neglected for correlation studies that imply low relative momentum, the total angular momentum $J$ equals the total spin $S$ and can take on values of $J = 2$ or $J=1$.
The $J = 2$ state cannot couple to the strangeness-rearrangement processes discussed above,
except through D-wave processes,
which are strongly suppressed.
For the $J = 1$ state only two limiting cases can be discussed in the absence of measurements of the $ \rm{p}\omm \rightarrow \Xi \La,\, \Xi\sig$ cross sections.

The first case assumes that the effect of the inelastic channels is negligible for both configurations and that the radial behaviour of the interaction is driven by elastic processes, following the lattice QCD potential (see Fig.~\ref{fig:pot}), for both the $J = 2$ and $J = 1$ channels. This results in a prediction, shown by the orange curve Fig.~\ref{fig:CF}, panel \textbf{b}, that is close to the data in the low \ks region.
The second limiting case assumes, following a previous prescription~\cite{Morita:2019rph}, that the $J = 1$ configuration is completely dominated by strangeness-rearrangement processes.
The obtained correlation function is shown by the blue curve in Fig.~\ref{fig:CF}, panel \textbf{b}.
This curve clearly deviates from the data.
Both theoretical calculations also include the effect of the Coulomb interaction and
they predict the existence of a \pOmbound bound state with a binding energy of \SI{2.5}{MeV},
which causes a depletion in the correlation function in the \ks region between \num{100} and \SI{300}{\MeVc}, because pairs that form a bound state are lost to the correlation yield.
The inset of Fig.~\ref{fig:CF} shows that in this \ks region the data are consistent with unity and do not follow either of the two theoretical predictions. 
 
At the moment, the lattice QCD predictions underestimate the data, but additional measurements are necessary to draw a firm conclusion on the existence of the bound state.
Measurements of $\La\mbox{--}\xim$ and $\Sigma^{0}\mbox{--}\xim$ correlations will verify experimentally the strength of possible non-elastic contributions.
Measurements of the \pOm correlation function in collision systems with slightly larger size (for example, \pPb collisions at the LHC~\cite{FemtopXi}) will clarify the possible presence of a depletion in \textit{C}(\ks). Indeed, the appearance of a depletion in the correlation function depends on the interplay between the average intra-particle distance (source size) and the scattering length
associated with the \pOm interaction~\cite{Morita:2019rph}.

\section*{Summary}
We have shown that the hyperon--proton interaction can be studied in unprecedented detail in \pp collisions at $\sqrt{s}=13$~Te\kern-.1emV\xspace at the LHC. We have demonstrated, in particular, that even the as-yet-unknown \pOm interaction can be investigated with excellent precision. 
The comparison of the measured correlation functions shows that the \pOm signal is up to a factor two larger than the \pXim signal. This reflects the large difference in the strong-attractive interaction predicted by the first-principle calculations by the HAL QCD collaboration. 
The correlation functions predicted by \HALQCD are in agreement with the measurements for the \pXim interaction.
For the \pOm interaction, the inelastic channels are not yet accounted for quantitatively within the lattice QCD calculations.
Additionally, the depletion in the correlation function that is visible in the calculations around \ks = \SI{150}{\MeVc}, owing to the presence of a \pOmbound bound state, is not observed in the measured correlation.
To draw quantitative conclusions concerning the existence of a \pOmbound bound state, we plan a direct measurement of the $\La\mbox{--}\xim$ and $\Sigma^{0}\mbox{--}\xim$ correlations and a study of the \pOm correlation in \pPb collisions in the near future.
Indeed,  with the upgraded ALICE apparatus~\cite{Abelevetal:2014cna} and the increased data sample size expected from the high luminosity phase of the LHC Run~3 and Run~4~\cite{YellowReport_better}, 
the missing interactions involving hyperons will be measured in \pp and \pPb collisions and this should enable us to answer the question about the existence of a new baryon--baryon bound state.
Since this method can be extended to almost any hadron--hadron pair, an unexpected avenue for high-precision tests of the strong interaction at the LHC has been opened.

\newenvironment{acknowledgement}{\relax}{\relax}
\begin{acknowledgement}
\section*{Acknowledgements}
We are grateful to T. Hatsuda and K. Sasaki from the HAL QCD Collaboration for their valuable suggestions and for providing the lattice QCD results regarding the \pXim and \pOm interactions. We are also grateful to A. Ohnishi, T. Hyodo, T. Iritani, Y. Kamiya and T. Sekihara for their suggestions and discussions.

We thank all the engineers and technicians of the LHC for their contributions to the construction of the experiment and the CERN accelerator teams for the performance of the LHC complex.
We acknowledge the resources and support provided by all Grid centres and the Worldwide LHC Computing Grid (WLCG) collaboration.
We acknowledge the following funding agencies for their support in building and running the ALICE detector:
A. I. Alikhanyan National Science Laboratory (Yerevan Physics Institute) Foundation (ANSL), State Committee of Science and World Federation of Scientists (WFS), Armenia;
Austrian Academy of Sciences, Austrian Science Fund (FWF): [M 2467-N36] and Nationalstiftung f\"{u}r Forschung, Technologie und Entwicklung, Austria;
Ministry of Communications and High Technologies, National Nuclear Research Center, Azerbaijan;
Conselho Nacional de Desenvolvimento Cient\'{\i}fico e Tecnol\'{o}gico (CNPq), Financiadora de Estudos e Projetos (Finep), Funda\c{c}\~{a}o de Amparo \`{a} Pesquisa do Estado de S\~{a}o Paulo (FAPESP) and Universidade Federal do Rio Grande do Sul (UFRGS), Brazil;
Ministry of Education of China (MOEC) , Ministry of Science \& Technology of China (MSTC) and National Natural Science Foundation of China (NSFC), China;
Ministry of Science and Education and Croatian Science Foundation, Croatia;
Centro de Aplicaciones Tecnol\'{o}gicas y Desarrollo Nuclear (CEADEN), Cubaenerg\'{\i}a, Cuba;
Ministry of Education, Youth and Sports of the Czech Republic, Czech Republic;
The Danish Council for Independent Research | Natural Sciences, the VILLUM FONDEN and Danish National Research Foundation (DNRF), Denmark;
Helsinki Institute of Physics (HIP), Finland;
Commissariat \`{a} l'Energie Atomique (CEA) and Institut National de Physique Nucl\'{e}aire et de Physique des Particules (IN2P3) and Centre National de la Recherche Scientifique (CNRS), France;
Bundesministerium f\"{u}r Bildung und Forschung (BMBF) and GSI Helmholtzzentrum f\"{u}r Schwerionenforschung GmbH, Germany;
General Secretariat for Research and Technology, Ministry of Education, Research and Religions, Greece;
National Research, Development and Innovation Office, Hungary;
Department of Atomic Energy Government of India (DAE), Department of Science and Technology, Government of India (DST), University Grants Commission, Government of India (UGC) and Council of Scientific and Industrial Research (CSIR), India;
Indonesian Institute of Science, Indonesia;
Centro Fermi - Museo Storico della Fisica e Centro Studi e Ricerche Enrico Fermi and Istituto Nazionale di Fisica Nucleare (INFN), Italy;
Institute for Innovative Science and Technology , Nagasaki Institute of Applied Science (IIST), Japanese Ministry of Education, Culture, Sports, Science and Technology (MEXT) and Japan Society for the Promotion of Science (JSPS) KAKENHI, Japan;
Consejo Nacional de Ciencia (CONACYT) y Tecnolog\'{i}a, through Fondo de Cooperaci\'{o}n Internacional en Ciencia y Tecnolog\'{i}a (FONCICYT) and Direcci\'{o}n General de Asuntos del Personal Academico (DGAPA), Mexico;
Nederlandse Organisatie voor Wetenschappelijk Onderzoek (NWO), Netherlands;
The Research Council of Norway, Norway;
Commission on Science and Technology for Sustainable Development in the South (COMSATS), Pakistan;
Pontificia Universidad Cat\'{o}lica del Per\'{u}, Peru;
Ministry of Science and Higher Education, National Science Centre and WUT ID-UB, Poland;
Korea Institute of Science and Technology Information and National Research Foundation of Korea (NRF), Republic of Korea;
Ministry of Education and Scientific Research, Institute of Atomic Physics and Ministry of Research and Innovation and Institute of Atomic Physics, Romania;
Joint Institute for Nuclear Research (JINR), Ministry of Education and Science of the Russian Federation, National Research Centre Kurchatov Institute, Russian Science Foundation and Russian Foundation for Basic Research, Russia;
Ministry of Education, Science, Research and Sport of the Slovak Republic, Slovakia;
National Research Foundation of South Africa, South Africa;
Swedish Research Council (VR) and Knut \& Alice Wallenberg Foundation (KAW), Sweden;
European Organization for Nuclear Research, Switzerland;
Suranaree University of Technology (SUT), National Science and Technology Development Agency (NSDTA) and Office of the Higher Education Commission under NRU project of Thailand, Thailand;
Turkish Atomic Energy Agency (TAEK), Turkey;
National Academy of  Sciences of Ukraine, Ukraine;
Science and Technology Facilities Council (STFC), United Kingdom;
National Science Foundation of the United States of America (NSF) and United States Department of Energy, Office of Nuclear Physics (DOE NP), United States of America. \end{acknowledgement}

\bibliographystyle{utphys}   \providecommand{\href}[2]{#2}\begingroup\raggedright\endgroup

\appendix
\section{Methods}
\label{sec:Methods}

\subsection*{Event Selection}
 
Events were recorded from inelastic \pp collisions by ALICE~\ref{Aamodt:2008zz} \ref{Abelev:2014ffa} at the LHC. 
A trigger that requires the total signal amplitude measured in the V0 detector~\ref{Abbas:2013taa} to exceed a certain threshold was used to select high-multiplicity (HM) events.
The V0 detector comprises two plastic scintillator arrays placed on both sides of the interaction point at pseudorapidities $\num{2.8}<\eta<\num{5.1}$ and $\num{-3.7}<\eta<\num{-1.7}$.
The pseudorapidity is defined as $\eta = -\textrm{ln} \left[ \textrm{tan} \left( \frac{\theta}{2}\right) \right]$, where $\theta$ is the polar angle of the particle with respect to the proton beam axis.

At $\sqrt{s}=13$~Te\kern-.1emV\xspace, in the HM events, 30 charged particles in the range $|\eta|<\num{0.5}$ are produced on average. This $\eta$ range corresponds to the region within \num{26} degrees of the transverse plane that is perpendicular to the beam axis.
The HM events are rare, constituting \SI{0.17}{\percent} of the \pp collisions that produce at least one charged particle in the pseudorapidity range $|\eta|<\num{1.0}$.
It was shown~\cite{ALICE:2017jyt} that HM events contain an enhanced yield of hyperons, which facilitates this analysis. The yield of \omm in HM events is at least a factor $5$ larger, on average, compared with that in total inelastic collisions~\ref{Acharya:2019kyh}.
A total of \events  HM events were analysed.
Additional details on the HM event selection can be found in a previous work~\cite{Acharya:2019kqn}.

\subsection*{Particle Tracking and Identification}

For the identification and momentum measurement of charged particles, the Inner Tracking System (ITS)~\ref{Aamodt:2010aa}, Time Projection Chamber (TPC)~\ref{Alme:2010ke}, and Time-Of-Flight (TOF)~\ref{Akindinov:2013tea} detectors of ALICE are used.
All three detectors are located inside a solenoid magnetic field (\SI{0.5}{T}) leading to a bending of the trajectories of charged particles.
The measurement of the curvature is used to reconstruct the particle momenta. 
Typical transverse momentum ($p_{\rm{T}}$) resolutions for protons, pions and kaons vary from about \SI{2}{\percent} for tracks with $\pt = \SI{10}{\GeVc}$ to below \SI{1}{\percent} for $\pt<\SI{1}{\GeVc}$.
The particle identity is determined by the energy lost per unit of track length inside the TPC detector and, in some cases, by the particle velocity measured in the TOF detector. 
Additional experimental details are discussed in a previous work~\ref{Abelev:2014ffa}.

Protons are selected within a transverse momentum range of $\SI{0.5}<\pt<\SI{4.05}{\GeVc}$.
They are identified requiring TPC information for candidate tracks with momentum $p<\SI{0.75}{\GeVc}$, whereas TPC and TOF information are both required for candidates with $p>\SI{0.75}{\GeVc}$. An incorrect identification of primary protons occurs in \SI{1}{\percent} of the cases, as evaluated by Monte Carlo simulations. 

Direct tracking and identification is not possible for \xim and \omm hyperons and their antiparticles, because they are unstable and decay as a result of the weak interaction within a few centimetres after their production.
The mean decay distances (evaluated as $c \times\tau$, where $\tau$ is the particle lifetime) of $\Xi^{-}(\xip)\rightarrow \La (\aLa) + \uppi^- (\uppi^+) $ and $\omm (\omp) \rightarrow \La (\aLa) + $K$^{-} ($K$^{+}) $ are \num{4.9} and \num{2.5}~\si{cm}, respectively~\ref{Tanabashi:2018oca}. 
Both decays are followed by a second decay of the unstable \La ($\aLa$) hyperon, $\La (\aLa) \rightarrow  \mathrm{p} (\bar{\mathrm{p}}) +\uppi^- (\uppi^+)$, with an average decay path of \SI{7.9}{cm}~(ref. \ref{Tanabashi:2018oca}). Consequently, pions ($\uppi^{\pm}$), kaons (K$^{\pm}$) and protons have to be detected and then combined to search for $\xim$($\xip$) and $\Omega^{-}$(\omp) candidates. Those secondary particles are identified by the TPC information in the case of the reconstruction of $\xim$($\xip$), and in the case of $\Omega^{-}$(\omp) it is additionally required that the secondary protons and kaons are identified in the TOF detector. To measure the $\xim$($\xip$) and $\Omega^{-}$(\omp) hyperons, the two successive 
weak decays need to be reconstructed. The reconstruction procedure is very similar for both hyperons and is described in detail previously~\ref{Aamodt:2011zza}.
Topological selections are performed to reduce the combinatorial background, evaluated via a fit to the invariant mass distribution.

\subsection*{Determination of the source size}

The widths of the Gaussian distributions constituting $S(\rs)$, and defining the source size, are calculated on the basis of the results of the analysis of the \pP correlation function in pp collisions at $\sqrt{s}=13$~Te\kern-.1emV\xspace by the ALICE collaboration \cite{Acharya:2020dfb}. Assuming a common source for all baryons, its size was studied as a function of the transverse mass of the baryon--baryon pair, $\smash{\mt = \left(\kt^{2} + m^{2}\right)^{1/2}}$, where $m$ is the average mass and $\smash[b]{\kt = \mid \bm{p}\bf{_{\textrm{T},~1}} + \bm{p}\bf{_{\textrm{T},~2}}\mid}$/2 is the transverse momentum of the pair. The source size decreases with increasing mass, which could reflect the collective evolution of the system.
The average transverse mass $\langle m_{\mathrm{T}}\rangle$ for the \pXim and \pOm pairs differ and are equal to \SI{1.9}{\GeVc} and \SI{2.2}{\GeVc}, respectively.
To determine the source sizes for these values, the measurement from \pP correlations (shown in figure 5 of ref.~\cite{Acharya:2020dfb}) is parameterized as $r_{\textrm{core}} = a \cdot \mt^{b}+c$, where $r_{\textrm{core}}$ denotes the width of the Gaussian distribution defining the source before taking into account the effect produced by short lived resonances.

In \pp collisions at \onethree, \xim and \omm baryons are produced mostly as primary particles, but about \num{2/3} of the protons originate from the decay of short-lived resonances with a lifetime of a few fm per \textit{c}.
As a result, the effective source size of both \pXim and \pOm is modified.
This effect is taken into account by folding the Gaussian source with an exponential distribution following the method outlined previously~\cite{Acharya:2020dfb}. 
The resulting source distribution can be characterized by an effective Gaussian source radius equal to \radiuspXi\si{\fm} for \pXim pairs and to \radiuspOmega\si{\fm} for \pOm pairs.
The quoted uncertainties correspond to variations of the parametrization of the \pP results according to their systematic and statistical uncertainties.

\subsection*{Corrections of the Correlation Function}

The correction factor $\xi(\ks)$ accounts for the normalization of the $\ks$ distribution of pairs from mixed-events, for effects produced by finite momentum resolution and for the influence of residual correlations.

The mixed-event distribution, $N_{\mathrm{mixed}}(\ks)$, has to be scaled down, because the number of pairs available from mixed events is much higher than the number of pairs produced in the same collision used in $N_{\mathrm{same}}(\ks)$. The normalization parameter $\mathcal{N}$ is chosen such that the mean value of the correlation function equals to unity in a region of \ks values where the effect of final-state interactions are negligible, $\num{500}<\ks<\SI{800}{\MeVc}$.

The finite experimental momentum resolution modifies the measured correlation functions at most by \SI{8}{\percent} at low \ks. A correction for this effect is applied. Resolution effects due to the merging of tracks that are very close to each other were evaluated and found to be negligible.

The two measured correlation functions are dominated by the contribution of the interaction between \pXim and \pOm pairs. Nevertheless, other contributions also influence the measured correlation function. They originate either from incorrectly identified particles or from particles stemming from other weak decays (such as protons from $\Lambda \rightarrow \mathrm{p}+\uppi^-$ decays) combined with primary particles.
Because weak decays occur typically some centimetres away from the collision vertex, there is no final-state interaction between their decay products and the primary particles of interest.
Hence, the resulting correlation function either will be completely flat or will carry the residual signature of the interaction between the particles before the decay. A method to determine the exact shape and relative yields of the residual correlations has been previously developed~\cite{FemtoRun1}\ref{PhysRevC.89.054916}, and it is used in this analysis. Such contributions are subtracted from the measured \pXim and \pOm correlations to obtain the genuine correlation functions. The residual correlation stemming from misidentification is evaluated experimentally~\cite{FemtopXi} and its contribution is also subtracted from the measured correlation function. 

The systematic uncertainties associated with the genuine correlation function arise from the following sources: (i) the selection of the proton, \xim($\xip)$ and \omm(\omp), (ii) the normalization of the mixed-event distributions, (iii) uncertainties on the residual contributions, and (iv) uncertainties due to the finite momentum resolution.
To evaluate the associated systematic uncertainties: (i) all single-particle and topological selection criteria are varied with respect to their default values and the analysis is repeated for 50 different random combinations of such selection criteria so that the maximum change introduced in the number of \pXim and \pOm pairs is 25\% and the changes in the purity of protons, \xim($\xip)$ and \omm(\omp) are kept below 3\%; (ii) the \ks-normalization range of the mixed-events is varied, and a linear function of \ks is also used for an alternative normalization which results in an asymmetric uncertainty; (iii) the shape of the residual correlations and its relative contribution are altered; and (iv) the momentum resolution and the used correction method are changed. The total systematic uncertainties associated with the genuine correlation function are maximal at low \ks, reaching a value of \SI{9}{\percent} and \SI{8}{\percent} for \pXim and \pOm, respectively. 

\subsection*{HAL QCD potentials}

Results from calculations by the HAL QCD Collaboration for the \pXim \cite{Sasaki:2019qnh} and \pOm \cite{Iritani:2018sra} interactions are shown in Figs. \ref{fig:CF},\ref{fig:pot}. Such interactions were studied via (2+1)-flavor lattice QCD simulations with nearly physical quark masses ($\textit{m}_{\pi}$ = 146 MeV/\textit{c}$^2$).

In Fig. \ref{fig:pot}, the \pXim and \pOm potentials are shown for calculations with \textit{t}/\textit{a} = 12, with \textit{t} the Euclidean time and \textit{a} the lattice spacing of the calculations.
The HAL QCD Collaboration provided 23 and 20 sets of parameters for the description of the shape of the \pXim and \pOm potentials, respectively.
Such parametrizations result from applying the jackknife method, which takes into account the statistical uncertainty of the calculations.
The width of the curves in Fig.~\ref{fig:pot} corresponds to the maximum variations observed in the potential shape by using the different sets of parameters.

To obtain the correlation functions shown in Fig. \ref{fig:CF} we consider the calculations with \textit{t}/\textit{a} = 12, both for \pXim and \pOm.
The statistical uncertainty of the calculations is evaluated using the jackknife variations, and a systematic uncertainty is added in quadrature evaluated by considering calculations with \textit{t}/\textit{a} = 11 and \textit{t}/\textit{a} = 13.

\subsection*{References} 
\begin{enumerate}[label={[\arabic*]}]
\setcounter{enumi}{49}

\item \label{Aamodt:2008zz} {\bfseries ALICE} Collaboration, K.~Aamodt {\em et~al.}, ``{The ALICE
  experiment at the CERN LHC}'',
  \href{http://dx.doi.org/10.1088/1748-0221/3/08/S08002}{{\em JINST} {\bfseries
  3} (2008) S08002}.

\item \label{Abelev:2014ffa} {\bfseries ALICE} Collaboration, B.~Abelev {\em et~al.}, ``{Performance of the
  ALICE Experiment at the CERN LHC}'',
  \href{http://dx.doi.org/10.1142/S0217751X14300440}{{\em Int. J. Mod. Phys.}
  {\bfseries A29} (2014) 1430044}.

\item \label{Abbas:2013taa} {\bfseries ALICE} Collaboration, E.~Abbas {\em et~al.}, ``{Performance of the
  ALICE VZERO system}'',
  \href{http://dx.doi.org/10.1088/1748-0221/8/10/P10016}{{\em JINST} {\bfseries
  8} (2013) P10016}.

\item \label{Acharya:2019kyh} {\bfseries ALICE} Collaboration, S.~Acharya {\em et~al.}, ``{Multiplicity
  dependence of (multi-)strange hadron production in proton-proton collisions
  at $\sqrt{s}$ = 13 TeV}'',
  \href{http://dx.doi.org/10.1140/epjc/s10052-020-7673-8}{{\em Eur. Phys. J. C}
  {\bfseries 80} (2020) 167}.

\item \label{Aamodt:2010aa} {\bfseries ALICE} Collaboration, K.~Aamodt {\em et~al.}, ``{Alignment of the
  ALICE Inner Tracking System with cosmic-ray tracks}'',
  \href{http://dx.doi.org/10.1088/1748-0221/5/03/P03003}{{\em JINST} {\bfseries
  5} (2010) P03003}.

\item \label{Alme:2010ke} J.~Alme {\em et~al.}, ``{The ALICE TPC, a large 3-dimensional tracking device
  with fast readout for ultra-high multiplicity events}'',
  \href{http://dx.doi.org/10.1016/j.nima.2010.04.042}{{\em Nucl. Instrum. Meth.
  A} {\bfseries 622} (2010) 316--367}.

\item \label{Akindinov:2013tea} A.~Akindinov {\em et~al.}, ``{Performance of the ALICE Time-Of-Flight detector
  at the LHC}'', \href{http://dx.doi.org/10.1140/epjp/i2013-13044-x}{{\em Eur.
  Phys. J. Plus} {\bfseries 128} (2013) 44}.

\item \label{Tanabashi:2018oca} {\bfseries Particle Data Group} Collaboration, M.~Tanabashi {\em et~al.},
  ``{Review of Particle Physics}'',
\href{http://dx.doi.org/10.1103/PhysRevD.98.030001}{{\em Phys. Rev.} {\bfseries
  D98} (2018) 030001}.

\item \label{Aamodt:2011zza} {\bfseries ALICE} Collaboration, K.~Aamodt {\em et~al.}, ``{Strange particle
  production in proton-proton collisions at sqrt(s) = 0.9 TeV with ALICE at the
  LHC}'', \href{http://dx.doi.org/10.1140/epjc/s10052-011-1594-5}{{\em Eur.
  Phys. J. C} {\bfseries 71} (2011) 1594}.

\item \label{PhysRevC.89.054916} A.~Kisiel, H.~Zbroszczyk, and M.~Szyma\ifmmode~\acute{n}\else \'{n}\fi{}ski, ``Extracting baryon-antibaryon strong-interaction potentials from $p\overline{\ensuremath{\Lambda}}$ femtoscopic correlation functions'', \href{http://dx.doi.org/10.1103/PhysRevC.89.054916}{{\em Phys. Rev. C} {\bfseries 89} (2014) 054916}.
  
\end{enumerate}

 \newpage
\section{The ALICE Collaboration}
\label{app:collab}

\begingroup
\small
\begin{flushleft}
S.~Acharya\Irefn{org141}\And 
D.~Adamov\'{a}\Irefn{org95}\And 
A.~Adler\Irefn{org74}\And 
J.~Adolfsson\Irefn{org81}\And 
M.M.~Aggarwal\Irefn{org100}\And 
G.~Aglieri Rinella\Irefn{org34}\And 
M.~Agnello\Irefn{org30}\And 
N.~Agrawal\Irefn{org10}\textsuperscript{,}\Irefn{org54}\And 
Z.~Ahammed\Irefn{org141}\And 
S.~Ahmad\Irefn{org16}
% ...
\And
\renewcommand\labelenumi{\textsuperscript{\theenumi}~}

\section*{Affiliation notes}
\renewcommand\theenumi{\roman{enumi}}
\begin{Authlist}
\item \Adef{org*}Deceased
\item \Adef{orgI}Italian National Agency for New Technologies, Energy and Sustainable Economic Development (ENEA), Bologna, Italy
\item \Adef{orgII}Dipartimento DET del Politecnico di Torino, Turin, Italy
\item \Adef{orgIII}M.V. Lomonosov Moscow State University, D.V. Skobeltsyn Institute of Nuclear, Physics, Moscow, Russia
\item \Adef{orgIV}Department of Applied Physics, Aligarh Muslim University, Aligarh, India
\item \Adef{orgV}Institute of Theoretical Physics, University of Wroclaw, Poland
\end{Authlist}

\section*{Collaboration Institutes}
\renewcommand\theenumi{\arabic{enumi}~}
\begin{Authlist}
\item \Idef{org1}A.I. Alikhanyan National Science Laboratory (Yerevan Physics Institute) Foundation, Yerevan, Armenia
\item \Idef{org2}Bogolyubov Institute for Theoretical Physics, National Academy of Sciences of Ukraine, Kiev, Ukraine
\item \Idef{org3}Bose Institute, Department of Physics  and Centre for Astroparticle Physics and Space Science (CAPSS), Kolkata, India
\item \Idef{org4}Budker Institute for Nuclear Physics, Novosibirsk, Russia
\item \Idef{org5}California Polytechnic State University, San Luis Obispo, California, United States
\item \Idef{org6}Central China Normal University, Wuhan, China
\item \Idef{org7}Centre de Calcul de l'IN2P3, Villeurbanne, Lyon, France
\item \Idef{org8}Centro de Aplicaciones Tecnol\'{o}gicas y Desarrollo Nuclear (CEADEN), Havana, Cuba
\item \Idef{org9}Centro de Investigaci\'{o}n y de Estudios Avanzados (CINVESTAV), Mexico City and M\'{e}rida, Mexico
\item \Idef{org10}Centro Fermi - Museo Storico della Fisica e Centro Studi e Ricerche ``Enrico Fermi', Rome, Italy
\item \Idef{org11}Chicago State University, Chicago, Illinois, United States
\item \Idef{org12}China Institute of Atomic Energy, Beijing, China
\item \Idef{org13}Comenius University Bratislava, Faculty of Mathematics, Physics and Informatics, Bratislava, Slovakia
\item \Idef{org14}COMSATS University Islamabad, Islamabad, Pakistan
\item \Idef{org15}Creighton University, Omaha, Nebraska, United States
\item \Idef{org16}Department of Physics, Aligarh Muslim University, Aligarh, India
\item \Idef{org17}Department of Physics, Pusan National University, Pusan, Republic of Korea
\item \Idef{org18}Department of Physics, Sejong University, Seoul, Republic of Korea
\item \Idef{org19}Department of Physics, University of California, Berkeley, California, United States
\item \Idef{org20}Department of Physics, University of Oslo, Oslo, Norway
\item \Idef{org21}Department of Physics and Technology, University of Bergen, Bergen, Norway
\item \Idef{org22}Dipartimento di Fisica dell'Universit\`{a} 'La Sapienza' and Sezione INFN, Rome, Italy
\item \Idef{org23}Dipartimento di Fisica dell'Universit\`{a} and Sezione INFN, Cagliari, Italy
\item \Idef{org24}Dipartimento di Fisica dell'Universit\`{a} and Sezione INFN, Trieste, Italy
\item \Idef{org25}Dipartimento di Fisica dell'Universit\`{a} and Sezione INFN, Turin, Italy
\item \Idef{org26}Dipartimento di Fisica e Astronomia dell'Universit\`{a} and Sezione INFN, Bologna, Italy
\item \Idef{org27}Dipartimento di Fisica e Astronomia dell'Universit\`{a} and Sezione INFN, Catania, Italy
\item \Idef{org28}Dipartimento di Fisica e Astronomia dell'Universit\`{a} and Sezione INFN, Padova, Italy
\item \Idef{org29}Dipartimento di Fisica `E.R.~Caianiello' dell'Universit\`{a} and Gruppo Collegato INFN, Salerno, Italy
\item \Idef{org30}Dipartimento DISAT del Politecnico and Sezione INFN, Turin, Italy
\item \Idef{org31}Dipartimento di Scienze e Innovazione Tecnologica dell'Universit\`{a} del Piemonte Orientale and INFN Sezione di Torino, Alessandria, Italy
\item \Idef{org32}Dipartimento di Scienze MIFT, Universit\`{a} di Messina, Messina, Italy
\item \Idef{org33}Dipartimento Interateneo di Fisica `M.~Merlin' and Sezione INFN, Bari, Italy
\item \Idef{org34}European Organization for Nuclear Research (CERN), Geneva, Switzerland
\item \Idef{org35}Faculty of Electrical Engineering, Mechanical Engineering and Naval Architecture, University of Split, Split, Croatia
\item \Idef{org36}Faculty of Engineering and Science, Western Norway University of Applied Sciences, Bergen, Norway
\item \Idef{org37}Faculty of Nuclear Sciences and Physical Engineering, Czech Technical University in Prague, Prague, Czech Republic
\item \Idef{org38}Faculty of Science, P.J.~\v{S}af\'{a}rik University, Ko\v{s}ice, Slovakia
\item \Idef{org39}Frankfurt Institute for Advanced Studies, Johann Wolfgang Goethe-Universit\"{a}t Frankfurt, Frankfurt, Germany
\item \Idef{org40}Fudan University, Shanghai, China
\item \Idef{org41}Gangneung-Wonju National University, Gangneung, Republic of Korea
\item \Idef{org42}Gauhati University, Department of Physics, Guwahati, India
\item \Idef{org43}Helmholtz-Institut f\"{u}r Strahlen- und Kernphysik, Rheinische Friedrich-Wilhelms-Universit\"{a}t Bonn, Bonn, Germany
\item \Idef{org44}Helsinki Institute of Physics (HIP), Helsinki, Finland
\item \Idef{org45}High Energy Physics Group,  Universidad Aut\'{o}noma de Puebla, Puebla, Mexico
\item \Idef{org46}Hiroshima University, Hiroshima, Japan
\item \Idef{org47}Hochschule Worms, Zentrum  f\"{u}r Technologietransfer und Telekommunikation (ZTT), Worms, Germany
\item \Idef{org48}Horia Hulubei National Institute of Physics and Nuclear Engineering, Bucharest, Romania
\item \Idef{org49}Indian Institute of Technology Bombay (IIT), Mumbai, India
\item \Idef{org50}Indian Institute of Technology Indore, Indore, India
\item \Idef{org51}Indonesian Institute of Sciences, Jakarta, Indonesia
\item \Idef{org52}INFN, Laboratori Nazionali di Frascati, Frascati, Italy
\item \Idef{org53}INFN, Sezione di Bari, Bari, Italy
\item \Idef{org54}INFN, Sezione di Bologna, Bologna, Italy
\item \Idef{org55}INFN, Sezione di Cagliari, Cagliari, Italy
\item \Idef{org56}INFN, Sezione di Catania, Catania, Italy
\item \Idef{org57}INFN, Sezione di Padova, Padova, Italy
\item \Idef{org58}INFN, Sezione di Roma, Rome, Italy
\item \Idef{org59}INFN, Sezione di Torino, Turin, Italy
\item \Idef{org60}INFN, Sezione di Trieste, Trieste, Italy
\item \Idef{org61}Inha University, Incheon, Republic of Korea
\item \Idef{org62}Institute for Nuclear Research, Academy of Sciences, Moscow, Russia
\item \Idef{org63}Institute for Subatomic Physics, Utrecht University/Nikhef, Utrecht, Netherlands
\item \Idef{org64}Institute of Experimental Physics, Slovak Academy of Sciences, Ko\v{s}ice, Slovakia
\item \Idef{org65}Institute of Physics, Homi Bhabha National Institute, Bhubaneswar, India
\item \Idef{org66}Institute of Physics of the Czech Academy of Sciences, Prague, Czech Republic
\item \Idef{org67}Institute of Space Science (ISS), Bucharest, Romania
\item \Idef{org68}Institut f\"{u}r Kernphysik, Johann Wolfgang Goethe-Universit\"{a}t Frankfurt, Frankfurt, Germany
\item \Idef{org69}Instituto de Ciencias Nucleares, Universidad Nacional Aut\'{o}noma de M\'{e}xico, Mexico City, Mexico
\item \Idef{org70}Instituto de F\'{i}sica, Universidade Federal do Rio Grande do Sul (UFRGS), Porto Alegre, Brazil
\item \Idef{org71}Instituto de F\'{\i}sica, Universidad Nacional Aut\'{o}noma de M\'{e}xico, Mexico City, Mexico
\item \Idef{org72}iThemba LABS, National Research Foundation, Somerset West, South Africa
\item \Idef{org73}Jeonbuk National University, Jeonju, Republic of Korea
\item \Idef{org74}Johann-Wolfgang-Goethe Universit\"{a}t Frankfurt Institut f\"{u}r Informatik, Fachbereich Informatik und Mathematik, Frankfurt, Germany
\item \Idef{org75}Joint Institute for Nuclear Research (JINR), Dubna, Russia
\item \Idef{org76}Korea Institute of Science and Technology Information, Daejeon, Republic of Korea
\item \Idef{org77}KTO Karatay University, Konya, Turkey
\item \Idef{org78}Laboratoire de Physique des 2 Infinis, Ir\`{e}ne Joliot-Curie, Orsay, France
\item \Idef{org79}Laboratoire de Physique Subatomique et de Cosmologie, Universit\'{e} Grenoble-Alpes, CNRS-IN2P3, Grenoble, France
\item \Idef{org80}Lawrence Berkeley National Laboratory, Berkeley, California, United States
\item \Idef{org81}Lund University Department of Physics, Division of Particle Physics, Lund, Sweden
\item \Idef{org82}Nagasaki Institute of Applied Science, Nagasaki, Japan
\item \Idef{org83}Nara Women{'}s University (NWU), Nara, Japan
\item \Idef{org84}National and Kapodistrian University of Athens, School of Science, Department of Physics , Athens, Greece
\item \Idef{org85}National Centre for Nuclear Research, Warsaw, Poland
\item \Idef{org86}National Institute of Science Education and Research, Homi Bhabha National Institute, Jatni, India
\item \Idef{org87}National Nuclear Research Center, Baku, Azerbaijan
\item \Idef{org88}National Research Centre Kurchatov Institute, Moscow, Russia
\item \Idef{org89}Niels Bohr Institute, University of Copenhagen, Copenhagen, Denmark
\item \Idef{org90}Nikhef, National institute for subatomic physics, Amsterdam, Netherlands
\item \Idef{org91}NRC Kurchatov Institute IHEP, Protvino, Russia
\item \Idef{org92}NRC \guillemotleft Kurchatov\guillemotright~Institute - ITEP, Moscow, Russia
\item \Idef{org93}NRNU Moscow Engineering Physics Institute, Moscow, Russia
\item \Idef{org94}Nuclear Physics Group, STFC Daresbury Laboratory, Daresbury, United Kingdom
\item \Idef{org95}Nuclear Physics Institute of the Czech Academy of Sciences, \v{R}e\v{z} u Prahy, Czech Republic
\item \Idef{org96}Oak Ridge National Laboratory, Oak Ridge, Tennessee, United States
\item \Idef{org97}Ohio State University, Columbus, Ohio, United States
\item \Idef{org98}Petersburg Nuclear Physics Institute, Gatchina, Russia
\item \Idef{org99}Physics department, Faculty of science, University of Zagreb, Zagreb, Croatia
\item \Idef{org100}Physics Department, Panjab University, Chandigarh, India
\item \Idef{org101}Physics Department, University of Jammu, Jammu, India
\item \Idef{org102}Physics Department, University of Rajasthan, Jaipur, India
\item \Idef{org103}Physikalisches Institut, Eberhard-Karls-Universit\"{a}t T\"{u}bingen, T\"{u}bingen, Germany
\item \Idef{org104}Physikalisches Institut, Ruprecht-Karls-Universit\"{a}t Heidelberg, Heidelberg, Germany
\item \Idef{org105}Physik Department, Technische Universit\"{a}t M\"{u}nchen, Munich, Germany
\item \Idef{org106}Politecnico di Bari, Bari, Italy
\item \Idef{org107}Research Division and ExtreMe Matter Institute EMMI, GSI Helmholtzzentrum f\"ur Schwerionenforschung GmbH, Darmstadt, Germany
\item \Idef{org108}Rudjer Bo\v{s}kovi\'{c} Institute, Zagreb, Croatia
\item \Idef{org109}Russian Federal Nuclear Center (VNIIEF), Sarov, Russia
\item \Idef{org110}Saha Institute of Nuclear Physics, Homi Bhabha National Institute, Kolkata, India
\item \Idef{org111}School of Physics and Astronomy, University of Birmingham, Birmingham, United Kingdom
\item \Idef{org112}Secci\'{o}n F\'{\i}sica, Departamento de Ciencias, Pontificia Universidad Cat\'{o}lica del Per\'{u}, Lima, Peru
\item \Idef{org113}St. Petersburg State University, St. Petersburg, Russia
\item \Idef{org114}Stefan Meyer Institut f\"{u}r Subatomare Physik (SMI), Vienna, Austria
\item \Idef{org115}SUBATECH, IMT Atlantique, Universit\'{e} de Nantes, CNRS-IN2P3, Nantes, France
\item \Idef{org116}Suranaree University of Technology, Nakhon Ratchasima, Thailand
\item \Idef{org117}Technical University of Ko\v{s}ice, Ko\v{s}ice, Slovakia
\item \Idef{org118}The Henryk Niewodniczanski Institute of Nuclear Physics, Polish Academy of Sciences, Cracow, Poland
\item \Idef{org119}The University of Texas at Austin, Austin, Texas, United States
\item \Idef{org120}Universidad Aut\'{o}noma de Sinaloa, Culiac\'{a}n, Mexico
\item \Idef{org121}Universidade de S\~{a}o Paulo (USP), S\~{a}o Paulo, Brazil
\item \Idef{org122}Universidade Estadual de Campinas (UNICAMP), Campinas, Brazil
\item \Idef{org123}Universidade Federal do ABC, Santo Andre, Brazil
\item \Idef{org124}University of Cape Town, Cape Town, South Africa
\item \Idef{org125}University of Houston, Houston, Texas, United States
\item \Idef{org126}University of Jyv\"{a}skyl\"{a}, Jyv\"{a}skyl\"{a}, Finland
\item \Idef{org127}University of Liverpool, Liverpool, United Kingdom
\item \Idef{org128}University of Science and Technology of China, Hefei, China
\item \Idef{org129}University of South-Eastern Norway, Tonsberg, Norway
\item \Idef{org130}University of Tennessee, Knoxville, Tennessee, United States
\item \Idef{org131}University of the Witwatersrand, Johannesburg, South Africa
\item \Idef{org132}University of Tokyo, Tokyo, Japan
\item \Idef{org133}University of Tsukuba, Tsukuba, Japan
\item \Idef{org134}Universit\'{e} Clermont Auvergne, CNRS/IN2P3, LPC, Clermont-Ferrand, France
\item \Idef{org135}Universit\'{e} de Lyon, Universit\'{e} Lyon 1, CNRS/IN2P3, IPN-Lyon, Villeurbanne, Lyon, France
\item \Idef{org136}Universit\'{e} de Strasbourg, CNRS, IPHC UMR 7178, F-67000 Strasbourg, France, Strasbourg, France
\item \Idef{org137}Universit\'{e} Paris-Saclay Centre d'Etudes de Saclay (CEA), IRFU, D\'{e}partment de Physique Nucl\'{e}aire (DPhN), Saclay, France
\item \Idef{org138}Universit\`{a} degli Studi di Foggia, Foggia, Italy
\item \Idef{org139}Universit\`{a} degli Studi di Pavia, Pavia, Italy
\item \Idef{org140}Universit\`{a} di Brescia, Brescia, Italy
\item \Idef{org141}Variable Energy Cyclotron Centre, Homi Bhabha National Institute, Kolkata, India
\item \Idef{org142}Warsaw University of Technology, Warsaw, Poland
\item \Idef{org143}Wayne State University, Detroit, Michigan, United States
\item \Idef{org144}Westf\"{a}lische Wilhelms-Universit\"{a}t M\"{u}nster, Institut f\"{u}r Kernphysik, M\"{u}nster, Germany
\item \Idef{org145}Wigner Research Centre for Physics, Budapest, Hungary
\item \Idef{org146}Yale University, New Haven, Connecticut, United States
\item \Idef{org147}Yonsei University, Seoul, Republic of Korea
\end{Authlist}
\endgroup
 
 \end{document}